\documentclass[12pt,a4paper,oneside]{article}
\usepackage[top=1in,bottom=1in,left=.5in,right=.5in]{geometry}
\usepackage[margin=.5in,small]{caption}

\usepackage{authblk}
\usepackage{cite}
\usepackage{hyperref}
\usepackage{doi}
\usepackage{url}
\usepackage{amsmath,mathtools,gensymb,bigints}
\usepackage{epsfig}
\usepackage{graphics}
\usepackage{amsfonts, amssymb, upgreek}
\usepackage{textcomp}
\usepackage[english]{babel}
\usepackage{blindtext}
\usepackage{color}

\graphicspath{{img/}}

\providecommand{\abs}[1]{\vert #1\vert}

\newcommand{\tumble}{\mathtt{T}}
\newcommand{\run}{\mathtt{R}}
\newcommand{\FMSD}{F_\textsc{\tiny MSD}}
\newcommand{\TMSD}{T_\textsc{\tiny MSD}}
\newcommand{\RMSD}{R_\textsc{\tiny MSD}}

\newcommand{\const}{\mathop{}\!\mathrm{const}}
\newcommand{\upd}{\mathop{}\!\mathrm{d}}
\newcommand{\arsinh}{\mathop{}\!\mathrm{arsinh}}
\DeclareMathOperator{\arcsinh}{arcsinh}

\newcommand{\ex}{\mathrm{e}}

\newcommand{\Or}{\mathrm{O}}
\begin{document}

\title{Langevin equations for the run-and-tumble of swimming bacteria}

\author[1]{G. Fier}
\author[1,2]{D. Hansmann\thanks{David.Hansmann@conicet.gov.ar}}
\author[1,2]{R. C. Buceta\thanks{rbuceta@mdp.edu.ar}}

\affil[1]{Instituto de Investigaciones F\'{\i}sicas de Mar del Plata, UNMdP and CONICET}
\affil[2]{Departamento de F\'{\i}sica, FCEyN, Universidad Nacional de Mar del Plata}
\affil[{ }]{Funes 3350, B7602AYL Mar del Plata, Argentina}

\maketitle

\abstract{
The run and tumble motions of a swimming bacterium are well characterized by two stochastic variables: the speed $v(t)$ and the change of direction or deflection \mbox{$x(t)=\cos\varphi(t)$}, where $\varphi(t)$ is the turning angle at time $t$. Recently, we have introduced [Soft Matter {\bf 13}, 3385 (2017)] a single stochastic model for the deflection $x(t)$ of an {\sl E. coli} bacterium performing both types of movement in isotropic media without taxis, based on available experimental data. In this work we introduce Langevin equations for the variables $(v,x)$, which for particular values of a control parameter $\beta$ correspond to run and tumble motions, respectively. These Langevin equations have analytical solutions, which make it possible to calculate the statistical properties of both movements in detail. Assuming that the stochastic processes $x$ and $v$ are not independent during the tumble, we show that there are small displacements of the center of mass along the normal direction to the axis of the bacterial body, a consequence of the flagellar unbundling during the run-to-tumble transition. Regarding the tumble we show, by means of the directional correlation, that the process is not stationary for tumble-times of the order of experimentally measured characteristic tumble-time. The mean square displacement is studied in detail for both movements even in the non-stationary regime. We determine the diffusion and ballistic coefficients for tumble- and run-times, establishing their properties and relationships.
}



\section{Introduction\label{sec:intro}}

Currently, self-propelled microorganisms (SPM) without taxis (or directed motion) are studied as systems capable of remaining out of equilibrium, transiting between two or more metastable states. SPM under taxis (chemical, thermal, or electromagnetic, among others) can retain this behaviour even by moving in the direction of the gradient of the guide field \cite{Eisenbach2004}. Flagellated SPM (bacteria, algae, protozoa, sperm, etc.) have developed efficient mechanisms to move in bulk fluids or thin fluid layers on moist surfaces \cite{Pedley1992}. The basic movements of flagellated bacteria include translational and rotational degrees of freedom which may be restricted by constraints that are imposed by the medium geometry or by neighbouring congeners. In turn, the required resources (nutrients, temperature, oxygen, or other) facilitate the development and movement to explore and colonize the environment \cite{Purcell1977,Lauffenburger1991,Webre2003}. 

The locomotion mode of the motile flagellated bacteria (MFB) is referred to as swimming when  bacterium moves individually in bulk liquid, and is called swarming when the bacteria move collectively in a liquid thin film over a moist surface. Low density populations of MFB immersed in fluid media without constraints and low Reynolds number show movement patterns which depend on their species and strain, {\it e.g.} run-tumble, run-reverse or run-reverse-flick. MFB move rotating their helical flagella ({\it e.g. Escherichia coli} or {\sl Salmonella typhimurium}), which are jointed to a reversible rotary motor \cite{Berg1973}. A bacterium runs or turns when its flagellum or flagella (forming a bundle) spin with a fixed chirality. The {\it E. coli} motion is reduced to two consecutive steps called run and tumble. During the run, the flagella bundle of {\it E. coli} spins counterclockwise (CCW), viewed from behind. In contrast, during the slowdown (with reverse thrust) the flagella spin clockwise (CW) \cite{Macnab1977}. The change in the spinning direction (CCW to CW) unbundles the flagella \cite{Turner2000} rotating {\it E. coli} around its center of mass \cite{Darnton2007}. It is widely accepted that, during the tumble, the bacterium only changes the direction, leaving the center of mass without movement. However, unbundling flagella slightly moves the center of mass of the bacterium \cite{Turner2000}; this little studied aspect will be dealt with in this paper. After tumbling, the motor reversal (CW to CCW) forms a new flagella bundle that spins generating a drive in the new run direction. {\it E. coli} is the most studied MFB, in both its genomic and its internal biochemical processes \cite{Berg2004,Eisenbach2004}. Well-established experimental results concerning the run-and-tumble movement of {\it E. coli} allow further theoretical studies and conclusions can be extended to other MFB. 

\begin{figure}[h!]
\centering
\includegraphics[scale=0.25]{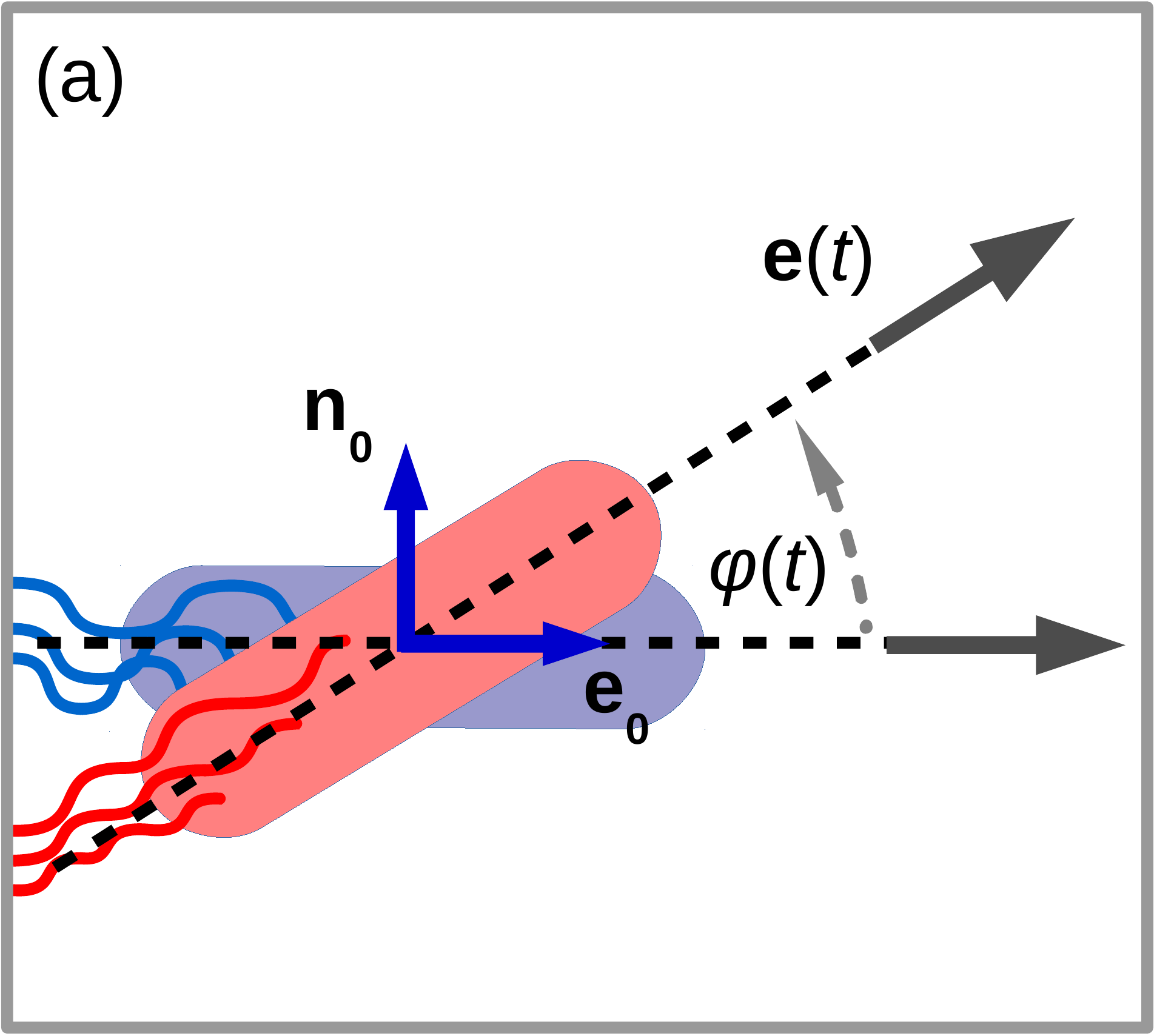}\hspace{1cm}
\includegraphics[scale=0.25]{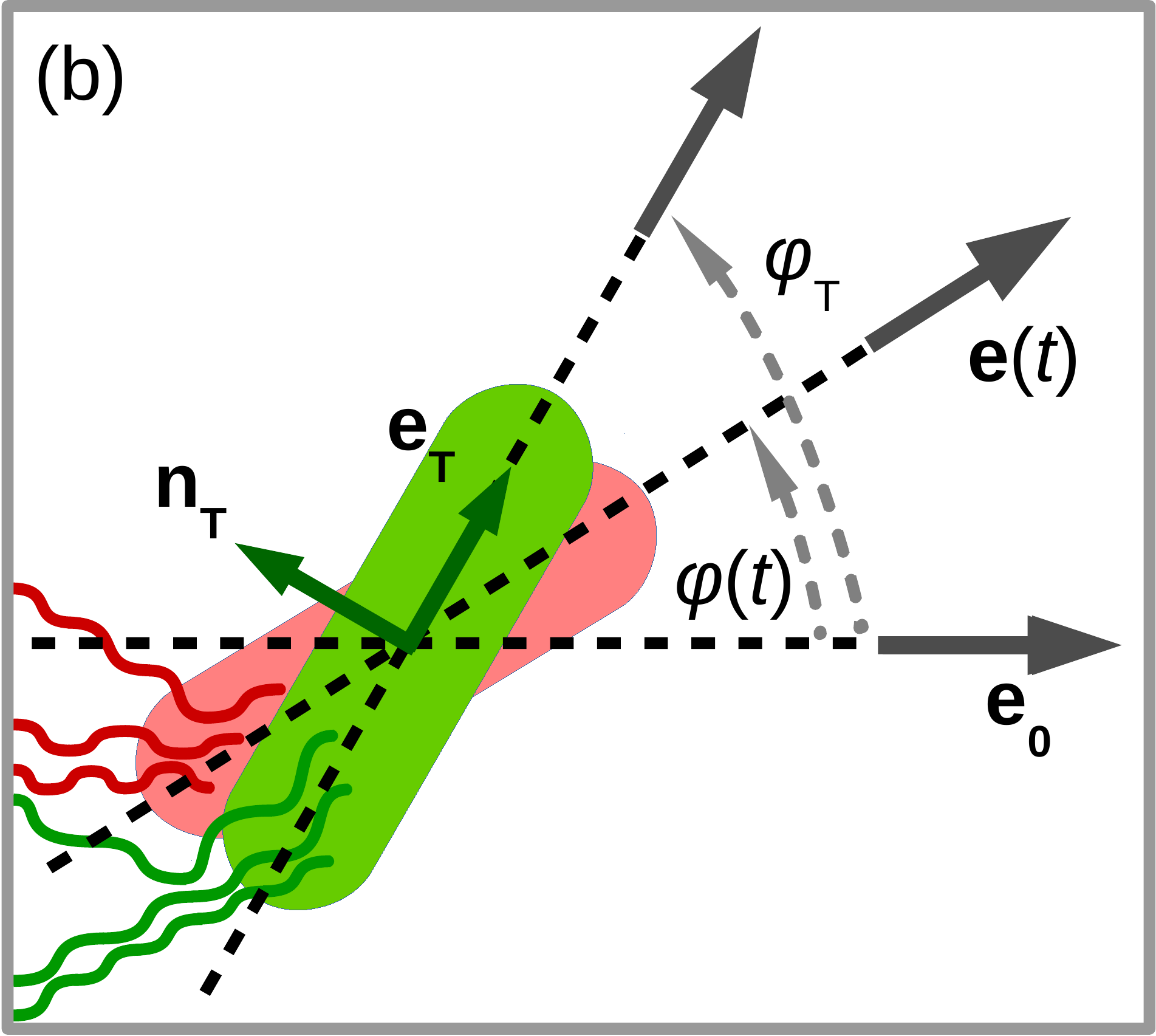}
\caption{(color online) Left plot: Shows the orientation of the bacterium at time $t$ in terms of the unit vector $\mathbf{e}(t)$, with $\varphi(t)$ being the turning angle respect to the incoming direction $\mathbf{e}_0=\mathbf{e}(t_0)$, for both tumble and run motions. Equation~(\ref{eq:e(t)}) gives us the vector $\mathbf{e}(t)$ in terms of the canonical basis $(\mathbf{e}_0,\mathbf{n}_0)$ and its normal vector is \mbox{$\mathbf{n}(t)=\sqrt{1-x^2}\,\mathbf{e}_0-x\,\mathbf{n}_0$\,}. Right plot: Shows the orientation of the bacterium at time $t$ in terms of the unit vector $\mathbf{e}(t)$, with \mbox{$\varphi(t)-\varphi_\tumble$} the angle respect to the outgoing direction $\mathbf{e}_\tumble=\mathbf{e}(t_\tumble)$, for the tumble motion. Equation~(\ref{eq:e(t)-approx}) gives us the vector $\mathbf{e}(t)$ in terms of the canonical basis $(\mathbf{e}_\tumble,\mathbf{n}_\tumble)$.  
\label{fig:base}}
\end{figure}

The path of a swimming bacterium consists of quasi-straight sections called runs, which are connected by tumbles or abrupt turns. This path is characterized at a time $t$ by its position $\mathbf{r}(t)$ in the three-dimensional reference frame. Each tumble motion is performed in the plane spanned by two consecutive runs ({\it i.e.} run-tumble-run), here called tumble-plane. Successive tumble-planes are connected to each other by a rotation. The velocity $\mathbf{v}(t)$ of the swimming bacterium on a tumble-plane can be specified by defining an intrinsic reference frame with two coordinate axes, where $(v_x,v_y)$ are Cartesian coordinates or $(v,\varphi)$ are polar coordinates. In the tumble-plane the velocity is
\begin{equation}
\mathbf{v}(t)=v(t)\,\mathbf{e}(t)\;,\label{eq:v}
\end{equation}
where $\mathbf{e}(t)$ is the heading unit vector, which sets the orientation (or polarity) of the bacterium, {\it e.g.} the {\it E. coli} orientation from the tail to the head. There is always a preferred direction of movement, which usually matches head-tail axis. However, the orientation does not always coincide with the direction of movement. The speed $v(t)$ can be positive or negative according to the bacteria moving forwards or backwards, respectively, {\it e.g.} running or tumbling {\it E. coli} have speeds which are greater than or equal to zero. The orientation, in terms of the deflection $x(t)=\cos[\varphi(t)]$, is 
\begin{equation}
\mathbf{e}(t)=x(t)\,\mathbf{e}_0+\sqrt{1-[x(t)]^2}\;\mathbf{n}_0\;,\label{eq:e(t)}
\end{equation}
with $(\mathbf{e}_0,\mathbf{n}_0)$ the canonical basis on the tumble-plane and where $\mathbf{e}_0=\mathbf{e}(t_0)$, taking $\varphi(t_0)=0$ (see left plot of Figure~(\ref{fig:base})). The movement of the SPMs is frequently studied as a the stochastic process described by stochastic differential equations (called Langevin equations). The Langevin equations can include terms of external and self-propelling forces, and noise (due to random forces or torques). Particularly, Langevin equations have been used to model experimental observations on the cells' motion \cite{Amselem2012, Selmeczi2005}. In order to find the Langevin equation for the velocity $\mathbf{v}$ it is useful to note that the acceleration, in terms of its tangential and normal components, is
\begin{equation}
\dot{\mathbf{v}}(t)=\dot{v}(t)\,\mathbf{e}(t)+\frac{v(t)\,\dot{x}(t)}{\sqrt{1-[x(t)]^2}}\;\mathbf{n}(t)\;,\label{eq:velocity}
\end{equation}
where $\mathbf{n}(t)$ is the normal vector to the trajectory (see Figure~\ref{fig:base} for details) and the overdot indicates derivative with respect to time. It is easy to show that the component of the normal acceleration is $a_\mathrm{n}=-v\,\dot{\varphi}$ and the angular velocity is $\mathbf{e}\wedge\dot{\mathbf{e}}=\dot{\varphi}\,\mathbf{u}$\,, where $\mathbf{u}=\mathbf{e}_0\wedge\mathbf{n}_0$ is the normal unit vector to the tumble-plane. In this work, we propose Langevin equations for the stochastic variables $x(t)$ and $v(t)$, which determine the velocity $\mathbf{v}(t)$ on a tumble-plane given by equation~(\ref{eq:v}). In such way, we can treat the run and tumble movements following two-dimensional classical approaches \cite{Schienbein1993}. The velocity $\mathbf{v}(t)$ of a bacterium is assumed to be a continuous-time stochastic process in 2-dimensions whose statistical properties will be studied in this paper. On this reference frame, in a lapse $[t_0,t]$ the displacement of the center of mass of a bacterium (for run or tumble motion) is $\mathbf{r}(t)-\mathbf{r}(t_0)=\int_{t_0}^t\mathbf{v}(t')\,\upd t'$. We show that the velocity correlation $\langle\mathbf{v}(t')\cdot\mathbf{v}(t'')\rangle$ and the mean-squared displacement (MSD) $\langle\abs{\mathbf{r}(t)-\mathbf{r}(t_0)}^2\rangle$ for the run and tumble movements can be conveniently described in this reference frame. 
The noise includes the effects of the collisions of the SPM, or the internal processes responsible for locomotion ({\it e.g}. flagellar motor), among others. The noise takes into account all the fast variables of the system which have very small characteristic-times compared to the time-scale of tumble or run processes. During the run bacteria move steadily forwards with noise fluctuations in their orientation and speed. On the contrary, during the tumble, the bacteria stop moving forwards and perform turning movements that include noise fluctuations in their orientation and speed. Throughout a century, great efforts have been made to develop a theoretical description that includes all the characteristics experimentally observed in the run and tumble movements. Initially, the movement of the SPMs was associated with the Brownian motion \cite{Selmeczi2007}; subsequently, the tendency to maintain the run direction led to the idea of persistent random walks \cite{Romanczuk2012}. Most of the existing models which describe the run and tumble motion using a Langevin equation for the speed $v(t)$ include a drift term capable of describing both movements in a steady state. Fier, Hansmann and Buceta (FHB) \cite{Fier2017} have recently derived Langevin equations for the orientation change $x(t)$ of the bacterium in the run and tumble movements which are able to describe  completely the well-known experimental results of Berg and Brown (BB) \cite{Berg1972}. Based on the same experimental data, six years ago Saragosti {\sl et al.}~\cite{Saragosti2012} proposed a pure rotational diffusion process in order to model the tumble motion. In contrast, the FHB model assumes that the tumble motion is the result of a stochastic process governed by rotational thrust and complemented by noise. In addition, the same Langevin equation is able to model the orientation changes of the bacteria during the run motion, varying the parameters of the FHB model to reproduce a Ornstein-Uhlenbeck (OU) process. The tumble processes are usually negligible when it is assumed that the tumble-time is one order of magnitude less than the runtime. FHB model has shown that the ratio between mean runtime and mean tumble-time allows to establish a single model for both movements. The changes of orientation or deflections $x(t)$ of the bacteria are fully described by the following Langevin equation 
\begin{equation}
\dot{x}=-\frac{\upd U}{\upd x}+\eta_{x}(t)\;,\label{eq:Langevin-deflection}
\end{equation} 
where $\eta_x(t)$ is Gaussian white noise with zero-mean and $U=U(x)$ is a phenomenological potential which has been derived by FHB from the measurements made by BB for the tumble-angle distribution $P(\varphi_\tumble)$ \cite{Berg1972}. Assuming that the probability density function (PDF) of tumble-angle deflection is $P(x_\tumble)=\mathcal{N}\,\ex^{-U(x_\tumble)}$, with $x_\tumble=\cos\varphi_\tumble$ and $\mathcal{N}$ the normalization constant, we proposed the potential 
\begin{equation}
U(x)=U_0-\rho\Bigl[x-\frac{\gamma}{\delta}\,\cosh(\delta x)\Bigl]\,\label{eq:potential}
\end{equation}
in reference \cite{Fier2017}, where $U_0$ is an adjustable constant and the constant parameters $\{\delta,\gamma,\rho\}$ are able to specify both movement states, {\sl i.e.} the deflection in the run and tumble. The three parameters are linked together by the phenomenological relationship
\begin{equation}
\rho^2\delta\sqrt{1+\gamma^2}= C = \const\;.\label{eq:phenom-rel}
\end{equation}
The values are $\rho(\delta_\tumble,\gamma_\tumble)=1$ for the tumble and $\rho(\delta_\run,\gamma_\run)=r>1$ for the run, where the sub-index $\tumble$ ($\run$) denotes the value that 
corresponds to the tumble (run) movement. The estimated values are $\delta_\tumble=9.062$ and $\gamma_\tumble=6.63\cdot 10^{-3}$ for the tumble motion, and $\delta_\run=4.71\cdot 10^{-2}$ and $\gamma_\run=4.98$ for the run motion \cite{Fier2017}. With this data, using equation~(\ref{eq:phenom-rel}), the constant is $C\approxeq\delta_\tumble$ and the calculated ratio $r\approxeq\sqrt{\delta_\tumble/(\delta_\run\gamma_\run)}\approxeq 6.21$ is very close to the experimental ratio for {\sl E. coli} measured by Berg and Brown $r_\mathrm{exp}\approx 6.14$ \cite{Berg1972}. 

Experimentally, it has been shown that the run- and tumble-times are random variables that follow an exponential distribution \cite{Berg1972}. Here, the run- or tumble-times correspond to the moments when the run or tumble movements stop, respectively. FHB have shown that the turning of the bacteria in the tumble motion is well characterized by a non-stationary stochastic process  \cite{Fier2017}. Also, FHB have shown that the turning in the run motion is characterized by an Ornstein-Uhlenbeck process. The FHB model shows that different turn movements of swimming {\sl E. coli} are characterized by a control parameter $\beta$ that takes values $\beta_\tumble\lessapprox 1$ (for tumble) or $\beta_\run\lessapprox 0$ (for run). Both values of $\beta$ are connected by a transition that passes through the critical value $\beta_\mathrm{c}=0$. The parameter $\beta$ is related to the parameters $\{\delta, \gamma, \rho\}$ of Langevin equation~(\ref{eq:Langevin-x}) by (see details in reference~\cite{Fier2017})
\begin{equation}
\beta(\gamma,\delta)=-\frac{4}{\gamma^2}\left(\frac{\gamma\,\ex^{-\delta}+1-\sqrt{1+\gamma^2}}{\gamma\,\ex^{-\delta}+1+\sqrt{1+\gamma^2}}\right)\,,
\end{equation}
which is used to estimate values for {\sl E. coli}: $\beta_\tumble\approx 0.965$ and $\beta_\run\approx -0.010$.

In the context of SPM modeling it is important to note that, besides minor similarities, there are big differences between colloidal systems and SPMs. Colloids are passive particles that perform Brownian motion in thermal equilibrium and whose mean velocity goes to zero for long times. In contrast, SPMs are particles that move actively in their environment and have a characteristic nonzero mean velocity for long times. Without additional external forces on the particle, the Langevin equation for the speed of the SPMs is
\begin{equation}
\dot{v}=-\lambda(\mathbf{r},\mathbf{v})\,v+\eta_{v}(t)\;,\label{eq:Langevin-velocity}
\end{equation}
where $\lambda$ is an effective friction coefficient that depends on velocity $\mathbf{v}$ in all cases. The noise $\eta_v(t)$ is Gaussian white with zero-mean. All models proposed for the coefficient of friction  $\lambda(\mathbf{v})$ have in common that, at high speeds, energy is dissipated with $\lambda(\mathbf{v})>0$ and, for low speeds, the internal energy is converted into active motion with $\lambda(\mathbf{v})<0$ (active friction) \cite{Schweitzer2003}. A well-known example of a velocity-dependent friction function that vanishes at $v = v_\mathrm{s}$, introduced by Schienbein and Gruler \cite{Schienbein1993}, is
\begin{equation}
\lambda(\mathbf{v})=\lambda_0\left(1-\frac{v_\mathrm{s}}{v}\right)\,,\label{eq:friction}
\end{equation} 
with $v> 0$, which allows to describe the active movement of different types of cells \cite{Gruler1994}. In this work we will give a very precise physical meaning at the steady speed $v_\mathrm{s}$ in terms of the values that $\beta$ takes in both run and tumble movements.

Here, we investigate a single stochastic model for the run and tumble motions of a swimming flagellated bacterium, starting from the Langevin equations for the speed $v(t)$ and deflection $x(t)$, and we study statistical properties of both movements. In Section~{\ref{sec:two}, we present the Langevin equations and their solutions {\it via} the Green function method. Statistical properties of the tumble motion are determinated in Section~\ref{sec:three}. First, we show that there is a small normal component of the average velocity to the axis of the bacterium body, which is a consequence of flagellar unbundling in the transition from CCW to CW. Then, we show that the directional correlation is non-stationary at tumble-times. Finally, in the same section, we show that the mean square displacement (MSD) is a relevant quantity in the tumble motion, even though the translation speeds are negligible, since it can explain the diffusion and ballistic behaviours at tumble times. In Section~{\ref{sec:four}, we determine the MSD of the run motion as a function of the initial conditions and the noise intensities for both the speed and the deflection, obtaining the properties of the diffusion and the ballistic movement at runtimes. Finally, we present the conclusions of this work and an outlook.

\section{Langevin equations and its solutions for the run and tumble motions\label{sec:two}}

The velocity $\mathbf{v}$ of a flagellated bacterium that performs run and tumble movements can be completely characterized by its speed $v$ and deflection $x$ respecting a reference frame, both as a function of time $t$. We propose Langevin equations for the stochastic variables $x$ and $v$. For deflection $x$ we use equation~(\ref{eq:Langevin-x}) in view of the fact that it reproduces the experimental results \cite{Fier2017}. For the speed $v$ we use a slightly generalized version of equation~(\ref{eq:Langevin-velocity}) referring to the results of previous studies \cite{Schweitzer2003}. 

Introducing the steady speed $v_\mathrm{s}$ as a function of the control parameter $\beta$ on the friction function $\lambda(\mathbf{v})$ given by equation~(\ref{eq:friction}), the Langevin equation~(\ref{eq:Langevin-velocity}) for the dimensionless speed $v=v(\beta,t)$ is
\begin{equation}
\dot{v}=-\lambda_0(\beta)\bigl[v-v_\mathrm{s}(\beta)\bigr]+\eta_{v}(t)\;,\label{eq:Langevin-v}
\end{equation}
where 
\begin{equation}
v_\mathrm{s}(\beta)=\frac{\beta_\tumble-\beta}{\beta_\tumble-\beta_\run}\,,\label{eq:vs-beta}
\end{equation}
with $\beta_\run\le\beta\le\beta_\tumble$\,, and $\eta_v(t)$ being Gaussian white noise with zero-mean. Using equation~(\ref{eq:vs-beta}) the steady speeds are \mbox{$v_\mathrm{s}=1$} for the run motion and \mbox{$v_\mathrm{s}=0$} for the tumble motion. It is easy to observe that the relaxation coefficient (or asymptotic value of the friction force \mbox{$\lambda(\mathbf{v})$} when \mbox{$v\gg v_\mathrm{s}$}) is 
\begin{equation}
\lambda_0(\beta)=\frac{1}{\tau(\beta)}
\end{equation}
where $\tau(\beta)$ is the characteristic time of the run or tumble processes. The complete solution of equation~(\ref{eq:Langevin-v}) for speed, with initial condition $v(t_0)=v_0$, is
\begin{equation}
v(\beta,t)=v_\mathrm{s}(\beta)+[v_0-v_\mathrm{s}(\beta)]\,G_v^{(\beta)}(t,t_0)+\int_{t_0}^t\eta_v(s)\,G_v^{(\beta)}(t,s)\,\upd s\;,\label{eq:sl-v}
\end{equation}
where the Green function, for both run and tumble motions, is
\begin{equation}
G_v^{(\beta)}(t,t')=\ex^{-\abs{t-t'}/\tau(\beta)}\;.
\end{equation}  
The Langevin equation~(\ref{eq:Langevin-deflection}) for the deflection $x=x(\beta,t)$ of a bacterium making run or tumble motions is
\begin{equation}
\dot{x}=\rho\,[1-\gamma\,\sinh(\delta x)]+\eta_x(t)\;,\label{eq:Langevin-x}
\end{equation} 
and its complete solution (see details in reference~\cite{Fier2017}) is
\begin{equation}
x(\beta,t)=x_\mathrm{s}(\beta)+[x_0-x_\mathrm{s}(\beta)]\,G_x^{(\beta)}(t,t_0)+\int_{t_0}^t\eta_x(s)\,G_x^{(\beta)}(t,s)\,\upd s\;,\label{eq:sl-x}
\end{equation}
where the Green function is
\begin{equation}
G_x^{(\beta)}(t,t')\simeq\left\lbrace
\begin{array}{ll}
\frac{\displaystyle\ln\bigl[1-\beta\,\ex^{-\max(t,t')/\tau(\beta)}\bigr]}{\displaystyle\ln\bigl[1-\beta\,\ex^{-\min(t,t')/\tau(\beta)}\bigr]}\hspace{2ex}&\mathrm{if}\;\beta=\beta_\tumble\vspace{1.5ex}\\
\;\ex^{-\abs{t-t'}/\tau(\beta)}&\mathrm{if}\;\beta=\beta_\run
\end{array}
\right.\;,
\end{equation}  
and the steady state solution is
\begin{equation}
x_\mathrm{s}(\beta)=\left\lbrace
\begin{array}{ll}
\frac{\displaystyle 1}{\displaystyle\delta_\tumble}\arsinh
\Bigl(\frac{\displaystyle 1}{\displaystyle\gamma_\tumble}\Bigr)\hspace{2ex}&\mathrm{if}\;
\beta=\beta_\tumble\vspace{.5ex}\\
1 &\mathrm{if}\;\beta=\beta_\run
\end{array}
\right.\;.\label{eq:sl-x_st}
\end{equation}  
The characteristic time $\tau=\tau(\beta)$ of the run or tumble motions is \cite{Fier2017}
\begin{equation}
\tau=\frac{1}{\rho\,\delta\sqrt{1+\gamma^2}}\;,
\end{equation}
or using the phenomenological relationship given by equation~(\ref{eq:phenom-rel}) we find \mbox{$\rho=C\tau$}, where the constant $C\approxeq \delta_\tumble$\,. We assume that $\eta_{u}(t)$ (where $u=x,v$) are Gaussian white noises with zero-means, {\sl i.e.} $\bigl\langle\eta_{u}(t)\bigr\rangle=0$\, and correlations 
\begin{equation}
\bigl\langle \eta_{u}(t)\,\eta_{u}(t')\bigr\rangle=2\,Q_{uu}\,\delta(t-t')\;.
\end{equation}
Usually, it is assumed that the stochastic processes $(x, v)$ are independent as a consequence of the noises $(\eta_x,\eta_v)$ being uncorrelated. However, this hypothesis is unrealistic for the tumble because here speed and direction of motion are not totally decoupled. This can be concluded from the calculations of the average velocity shown in Section~\ref{sec:three}. Consequently, it is appropriate to assume the noises are cross-correlated
\begin{equation}
\bigl\langle \eta_x(t)\,\eta_v(t')\bigr\rangle=2\,Q_{xv}\,\delta(t-t')\;,
\end{equation}
where the noise intensity $Q_{xv}$ is different from zero for the tumble motion and is equal to zero for the run motion, as we will show in this paper. 

\section{Statistical properties of the tumble motion\label{sec:three}}

In order to describe the statistical properties of the tumble motion it is practical to use the heading unit vector in the canonical basis $(\mathbf{e}_\tumble,\mathbf{n}_\tumble)$ of an intrinsic reference frame, which is defined by bacterial orientation at the end of the tumble movement (or outgoing direction), so that $\mathbf{e}_\tumble=\mathbf{e}(t_\tumble)$. The bacterial orientation at the outgoing direction and at the beginning (or incoming direction) of a tumble is related by \mbox{$\mathbf{e}_\tumble=x_\tumble\,\mathbf{e}_0+\sqrt{1-x^2_\tumble}\,\mathbf{n}_0$}\; and \mbox{$\mathbf{n}_\tumble=-\sqrt{1-x^2_\tumble}\,\mathbf{e}_0+x_\tumble\,\mathbf{n}_0$}, where \mbox{$x_\tumble=\cos\varphi_\tumble$} is the tumble deflection and $(\mathbf{e}_0,\mathbf{n}_0)$ is the canonical basis at the beginning of the tumble (see right plot of Figure~\ref{fig:base}). Expanding the unit vector $\mathbf{e}(t)$ (given by equation~(\ref{eq:e(t)})) in the canonical basis $(\mathbf{e}_\tumble,\mathbf{n}_\tumble)$ around $x=x_\tumble$ up to second order, we obtain
\begin{equation}
\mathbf{e}(t)\simeq\biggl\{1-\frac{[x(\beta_\tumble,t)-x_\tumble]^2}{2(1-x^2_\tumble)}\biggr\}\,\mathbf{e}_\tumble-\frac{[x(\beta_\tumble,t)-x_\tumble]}{\sqrt{1-x^2_\tumble}}\biggl\{1+\frac{x_\tumble\,[x(\beta_\tumble,t)-x_\tumble]}{2\,(1-x^2_\tumble)}\biggr\}\,\mathbf{n}_\tumble\;.\label{eq:e(t)-approx}
\end{equation} 
Taking equation~(\ref{eq:v}) into account, the lowest-order approximation of the average velocity, close to the tumble deflection, is
\begin{equation}
\bigl\langle\mathbf{v}(\beta_\tumble,t)\bigr\rangle\simeq\bigl\langle v(\beta_\tumble,t)\bigr\rangle\,\mathbf{e}_\tumble-\frac{1}{\sqrt{1-x^2_\tumble}}\,\bigl\langle v(\beta_\tumble,t)\,[x(\beta_\tumble,t)-x_\tumble]\,\bigr\rangle\,\mathbf{n}_\tumble\;,\label{eq:mean-v}
\end{equation}
where (taking $v_\mathrm{s}(\beta_\tumble)=0$)
\begin{eqnarray}
&&\langle v(\beta_\tumble,t)\rangle = v_0\,\,G_v^{(\beta_\tumble)}(t,t_0)\;,\\
&&\langle x(\beta_\tumble,t)\rangle = x_\mathrm{s}(\beta_\tumble)+[x_0-x_\mathrm{s}(\beta_\tumble)]\, G_x^{(\beta_\tumble)}(t,t_0)\;,
\end{eqnarray}
and 
\begin{equation}
\langle v(\beta_\tumble,t)[x(\beta_\tumble,t)-x_\tumble] \rangle = \langle v(\beta_\tumble,t)\rangle\,\bigl[\langle x(\beta_\tumble,t)\rangle-x_\tumble\bigr]+\;2\,Q_{xv}\int_{t_0}^t G_x^{(\beta_\tumble)}(t,s)\,G_v^{(\beta_\tumble)}(t,s)\,\upd s\;.\label{eq:appendix-1}
\end{equation}
By introducing the function $w(t)=\beta\,\ex^{-t/{\tau_\tumble}}$, taking $t_0=0$ and using equation~(\ref{eq:mean-v}) it can be shown that the normal component of the average velocity at the end of the tumble is (see details in Section A of Appendix)
\begin{eqnarray}
\langle\mathbf{v}(\beta_\tumble,t)\rangle\cdot\mathbf{n}_\tumble \!\!&=&\!\! -\frac{1}{\sqrt{1-x^2_\tumble}}\biggl\{\frac{1}{\beta_\tumble}\,v_0\,[x_\mathrm{s}(\beta_\tumble)-x_\tumble]\,w\nonumber\\ 
&&\hspace{1.7cm}+ \frac{1}{\beta_\tumble}\Bigl\{v_0\,[x_\mathrm{s}(\beta_\tumble)-x_\tumble]-2\,Q_{xv}\tau_\tumble\,\beta_\tumble\,I_{xv}(\beta_\tumble)\Bigr\}\,w\,\ln(1-w)\nonumber\\ 
&&\hspace{1.7cm}+\;2\,Q_{xv}\tau\,w\,\ln(1-w)\,I_{xv}(w)\biggr\}\;,\label{eq:v-normal-comp}
\end{eqnarray}
where
\begin{equation}
I_{xv}(u)=-\frac{1}{2\,u^2}+\frac{1}{2\,u}-\frac{\ln u}{12}-\frac{u}{24}-\frac{19\,u^2}{1440}-\cdots\;,\label{eq:appendix-2}
\end{equation}
with $0<u\le\beta\lessapprox 1$, is a Laurent series. The normal component is nonzero for all tumble-times $t_\tumble >0$. Particularly, taking into account that $w\,\ln(1-w)\,I_{xv}(w)\to\frac{1}{2}$ when $w\to 0$ (or $t\to +\infty$), it is easy to see that the normal component reaches a steady value (sv) for sufficiently long times ({\sl i.e.} $t\gg\tau_\tumble$) given by
\begin{equation}
\langle\mathbf{v}(\beta_\tumble,t)\rangle\cdot\mathbf{n}_\tumble\;\Bigl\rfloor_\mathrm{sv} = -\frac{Q_{xv}\,\tau_\tumble}{\sqrt{1-x^2_\tumble}}\;,
\end{equation}
with $Q_{xv}\neq 0$. Equation~(\ref{eq:v-normal-comp}) shows that, during the tumble, there is a small displacement of the center of mass in the normal direction to the axis of the bacterium, which is a consequence of assuming that the stochastic processes $x$ and $v$ are not independent. This result can be explained by the displacement of the center of mass as a consequence of the flagellar unbundling during the transition from CCW to CW. This conclusion is also valid when using the initial condition \mbox{$v_0=0$}, since in this case the mean velocity has only a normal component, which depends exclusively on the noise intensity $Q_{xv}$.

Another statistic observable of interest is the directional correlation function which, making use of equation~(\ref{eq:e(t)-approx}), is
\begin{equation}
\bigl\langle \mathbf{e}(t)\cdot\mathbf{e}(t')\bigr\rangle = \bigl\langle \cos[\theta(t,t')]\bigr\rangle \simeq 1-\frac{1}{2\,(1-x_\tumble^2)}\,\bigl\langle[x(t)-x(t')]^2\bigr\rangle+\Or(4)\;,\label{eq:dir-correlation}
\end{equation}
where $\theta(t,t')=\varphi(t)-\varphi(t')$ is the angle between the heading unit vectors at times $t$ and $t'$, which are close to the tumble-time $t_\tumble$. In the Appendix we show that the second-order term of the correlation given by equation~(\ref{eq:dir-correlation}) is
\begin{eqnarray}
\bigl\langle[x(t)-x(t')]^2\bigr\rangle \!\!&=&\!\! 2\,Q_{xx}\tau\;\biggl\{\biggl[\frac{[x_0-x_\mathrm{s}(\beta_\tumble)]^2}{2\,Q_{xx}\tau_\tumble\ln^2(1-\beta_\tumble)}-I_{xx}(\beta_\tumble)\biggl]\Bigl[\ln(1-w)-\ln(1-w')\Bigr]\label{eq:2-correlation}\\
&&\!\!+\;\Bigl[\ln^2(1-w)\,I_{xx}(w)-2\ln(1-w)\ln(1-w')\,I_{xx}(w_>)+\ln^2(1-w')\,I_{xx}(w')\Bigr]\biggr\},\nonumber
\end{eqnarray}
where $w=w(t)$, $w'=w(t')$, $w_>=\max(w,w')$, and
\begin{equation}
I_{xx}(u)=\frac{1}{2 u^2}-\frac{1}{u}-\frac{\ln u}{12}+\frac{u^2}{480}+\frac{u^3}{720}+\cdots\;,\label{eq:appendix-3}
\end{equation}
with $0<u\le\beta\lessapprox1$, is a Laurent serie (see Section B of Appendix). Note that the poles of $I_{xx}$ at $u = 0$ are removed from equation~(\ref{eq:2-correlation}) by the logarithmic functions. In particular, the term containing the second-order pole contributes to the stationary term (st) of equation~(\ref{eq:2-correlation}) 
\begin{equation}
\bigl\langle[x(t)-x(t')]^2\bigr\rangle_\mathrm{st}=2\,Q_{xx}\tau_\tumble\,\Bigl(1-\ex^{-\abs{t-t'}/\tau_\tumble}\Bigr)\;.
\end{equation}
It is easy to see that the remaining terms are non-stationary and dependent on $w$ and $w'$. Then the directional correlation up to the second order is
\begin{equation}
\bigl\langle \mathbf{e}(t)\cdot\mathbf{e}(t')\bigr\rangle \simeq 1-\,\frac{Q_{xx}\tau_\tumble}{(1-x_\tumble^2)}\,\Bigl(1-\ex^{-\abs{t-t'}/\tau_\tumble}\Bigr)+\mathcal{R}\bigl(t,t'\bigr)\;,\label{eq:dir-corr}
\end{equation}
where $\mathcal{R}$ includes non-stationary terms with the following properties: \mbox{$\lim_{z,z'\to +\infty}\mathcal{R}(z,z')=0$} and \mbox{$\mathcal{R}(z,z)=0$}. The directional correlation shows that the process is non-stationary since \mbox{$t=t_\tumble\approx\tau_\tumble$}, which is in agreement with the conclusion drawn from the covariance \cite{Fier2017}. Our result is quite different from the result corresponding to the diffusion of $\mathbf{e}(t)$ on the surface of a $d$-dimensional sphere of unit radius ($d\ge 2$), where the directional correlation is \mbox{$\bigl\langle \mathbf{e}(t)\cdot\mathbf{e}(t')\bigr\rangle=\ex^{-\abs{t-t'}/\tau_\tumble}$} \cite{Doi2001}. Even though the behaviours are similar, in our model the process is non-stationary and the correlation depends explicitly on the noise intensity $Q_{xx}$\,.

The mean square displacement (MSD) is defined by
\begin{equation}
\FMSD(t)\;\dot=\;\bigl\langle\abs{\mathbf{r}(t)-\mathbf{r}(t_0)}^2\bigr\rangle= \int_{t_0}^t\int_{t_0}^t\bigl\langle\mathbf{v}(t_1) \cdot\mathbf{v}(t_2)\bigr\rangle\,\upd t_1\,\upd t_2\;,\label{eq:MSD-def}
\end{equation}
where $\bigl\langle\mathbf{v}(t_1) \cdot\mathbf{v}(t_2)\bigr\rangle= \bigl\langle v(t_1)\,v(t_2)\,\cos[\theta(t_1,t_2)]\bigr\rangle$ is the velocity correlation function. If we assume that the stochastic processes $(x,v)$ are independent, which is equivalent to taking $Q_{xv} = 0$, the velocity correlation is $\bigl\langle\mathbf{v}(t_1) \cdot\mathbf{v}(t_2)\bigr\rangle= \bigl\langle v(t_1)\,v(t_2)\bigr\rangle\;\bigl\langle \cos[\theta(t_1,t_2)]\bigr\rangle$. Otherwise, if the processes are not independent ({\sl i.e.} $Q_{xv}\neq 0$), which is the case for the tumbling motion, the velocity correlation can be calculated from equations~(\ref{eq:v}) and (\ref{eq:e(t)-approx}) obtaining the following approximation
\begin{equation}
\bigl\langle\mathbf{v}(t_1)\cdot\mathbf{v}(t_2)\bigr\rangle\simeq \bigl\langle v(t_1)\,v(t_2)\bigr\rangle- \frac{1}{2(1-x_\tumble^2)}\,\bigl\langle v(t_1)\,v(t_2)\,[x(t_1)-x(t_2)]^2\bigr\rangle\;.
\end{equation}
\begin{figure}[h!]
\centering
\includegraphics[scale=0.55]{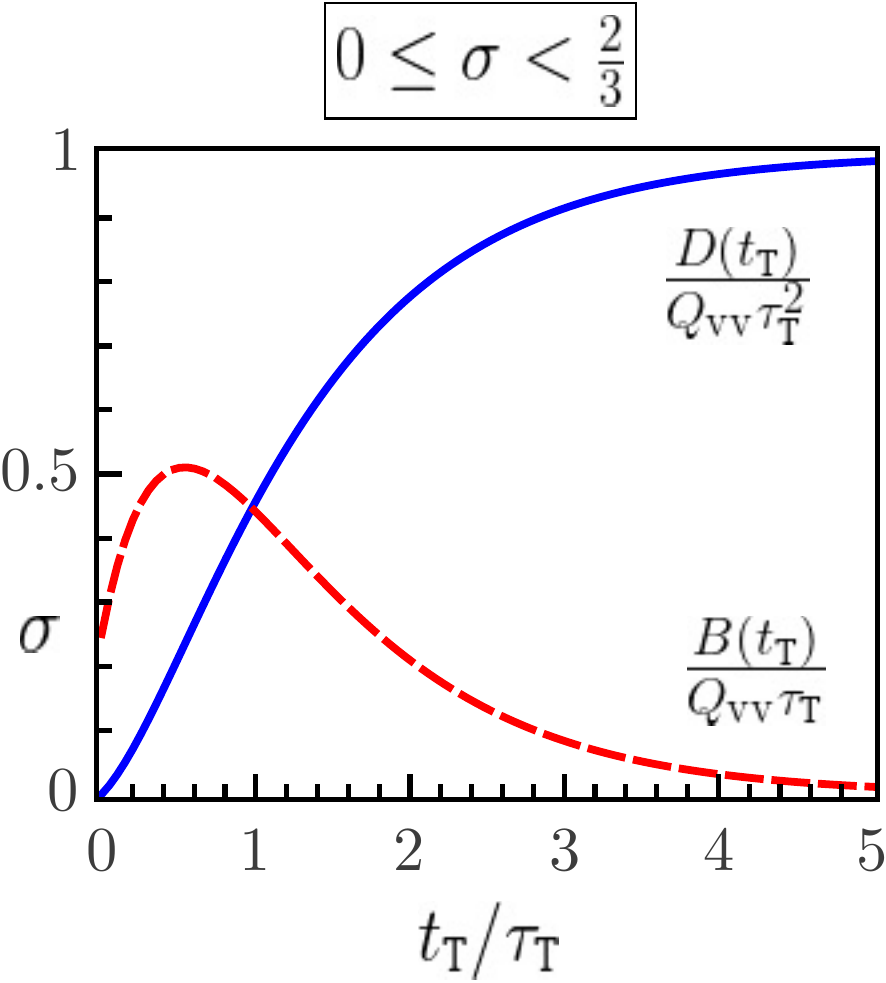}
\includegraphics[scale=0.5]{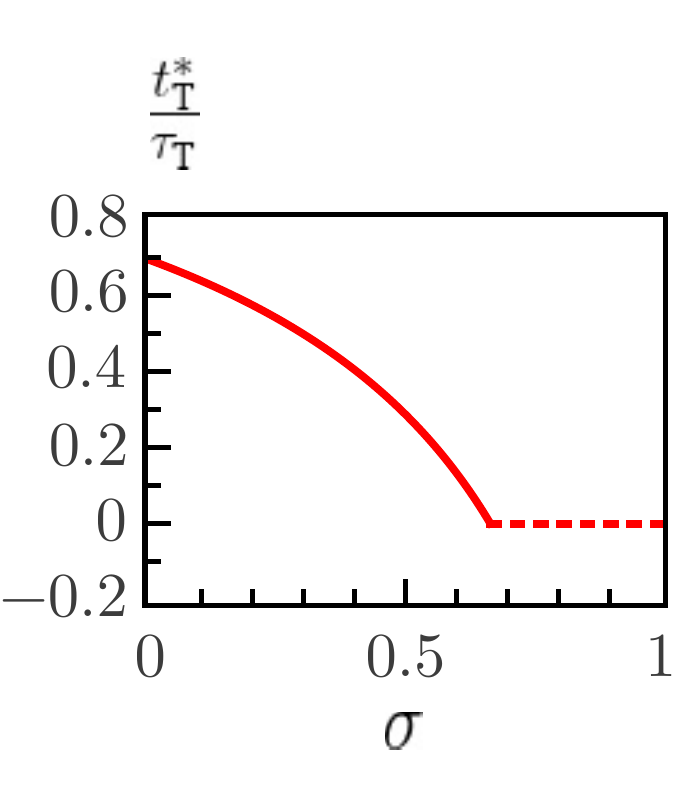}
\includegraphics[scale=0.55]{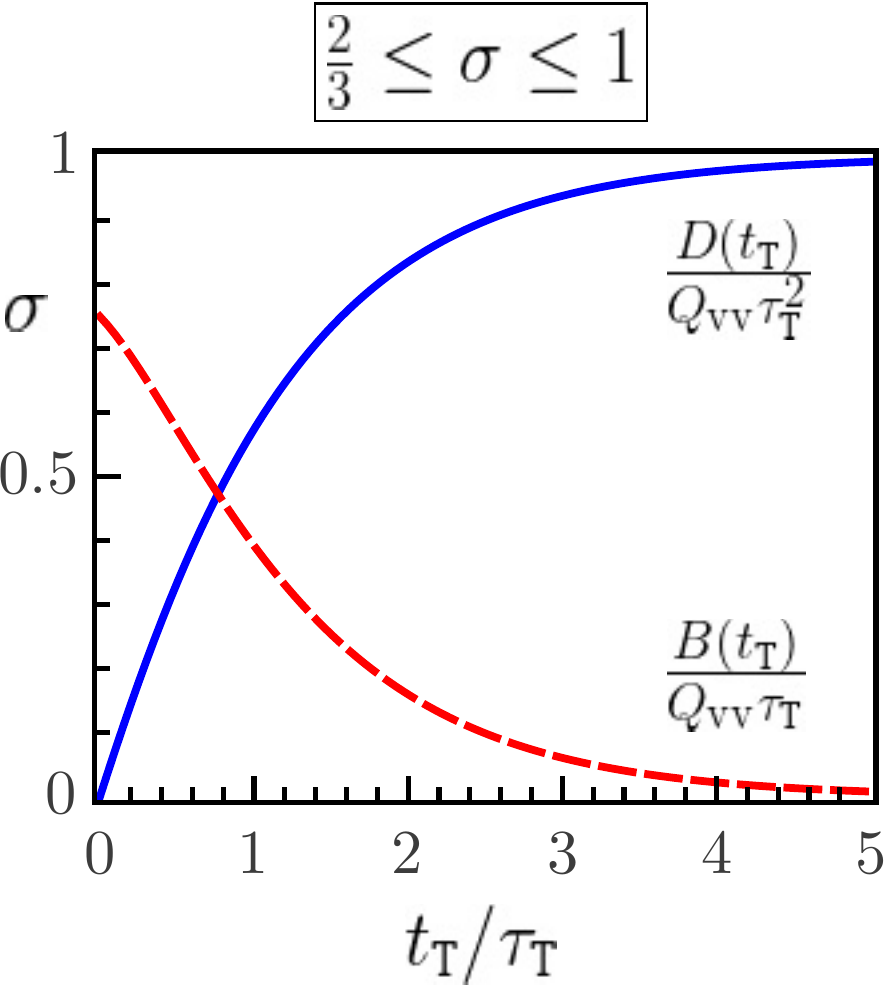}
\caption{(color online) Left and right plots: show the diffusion coefficient $D(t_\tumble)$ (blue continuous line) and square ballistic coefficient $D(t_\tumble)$ (red dashed line) as a function of the tumble-time $t_\tumble$ for two values of the parameter $\sigma$ (see equation~(\ref{eq:sigma})) included in the intervals mentioned in each plot. Both functions and time are dimensionless.  The ballistic contribution $B(t_\tumble)$ reaches a maximum for tumble-times \mbox{$t_\tumble^*=\tau_\tumble\,\ln[4(1-\sigma)/(2-\sigma)]$} if $0\le\sigma\le 2/3$ or \mbox{$t_\tumble^*=0$} if $2/3\le\sigma\le 1$.  Middle plot: shows the adimensionalized tumble-time $t^*_\tumble$ as a function of $\sigma$. Note that $t_\tumble^*/\tau_\tumble\le\ln 2\approxeq 0.693$, a value corresponding to the case $\sigma = 0$ (or zero initial velocity).
\label{fig:MSD-tumble}}
\end{figure}
By a standard calculation we found that the lowest order contribution ({\sl i.e.} the speed correlation function), with initial time $t_0=0$, is 
\begin{equation}
\bigl\langle v(t_1)\,v(t_2)\bigr\rangle = Q_\mathrm{vv}\tau_\tumble\,\Bigl[\ex^{-\abs{t_1-t_2}/\tau_\tumble}-(1-\sigma)\,\ex^{-(t_1+t_2)/\tau_\tumble}\Bigr]\;,\label{eq:v-corr}
\end{equation}
where 
\begin{equation}
\sigma=\frac{v_0^2}{Q_\mathrm{vv}\tau_\tumble}\;,\label{eq:sigma}
\end{equation}
which shows that taking $\sigma=1$ ensures the stationary behaviour \cite{Risken1989}, and where the mean square speed is $\langle v^2\rangle = v_0^2$\,. However, with this choice, the stationarity of the stochastic process $v$ is a particular property of the system. A suitable choice for the initial condition is $0\le\sigma <1$, so that it reproduces all possible non-stationarity situations. Then, for the tumble motion, the MSD contribution to lowest order is
\begin{equation}
\TMSD(t)=2\,Q_\mathrm{vv}\tau_\tumble^3\,\biggr[-\frac{3-\sigma}{2}+\frac{t}{\tau_\tumble}+(2-\sigma)\,\,\ex^{-t/\tau_\tumble}-\frac{1}{2}\,(1-\sigma)\,\,\ex^{-2\,t/\tau_\tumble}\biggr]\;.
\end{equation}
The usual asymptotic analysis to determine the diffusion constant is meaningless if the bacterium stops its turning motion at tumble-times which are of the order of the characteristic tumble-time. For this reason, we calculate its Taylor series at times close to the tumble-time ({\sl i.e.} $t\lessapprox t_\tumble$) to study the behaviour of the MSD 
\begin{equation}
\TMSD(t)=\TMSD(t_\tumble)+2\,D(t_\tumble)\,(t-t_\tumble)+B(t_\tumble)\,(t-t_\tumble)^2+\cdots\;,\label{eq:MSD-Taylor}
\end{equation}
where 
\begin{equation}
D(t_\tumble)=\frac{1}{2}\;\frac{\upd\TMSD}{\upd\,t}\biggr\rfloor_{t=t_\tumble}=Q_\mathrm{vv}\tau_\tumble^2\;\Bigr[1-(2-\sigma)\,\ex^{-t_\tumble/\tau_\tumble}+(1-\sigma)\,\ex^{-2\,t_\tumble/\tau_\tumble}\Bigr]\;\label{eq:diffusion-coef-tumble}
\end{equation}
is the diffusion coefficient at the tumble-time and 
\begin{equation}
B(t_\tumble)=\frac{1}{2}\;\frac{\upd^2 \TMSD}{\upd\,t^2}\biggr\rfloor_{t=t_\tumble}=Q_\mathrm{vv}\tau_\tumble\;\Bigr[(2-\sigma)-2\,(1-\sigma)\,\ex^{-t_\tumble/\tau_\tumble}\Bigr]\,\ex^{-t_\tumble/\tau_\tumble}\;\label{eq:ballistic-coef-tumble}
\end{equation}
is the square ballistic coefficient at the tumble-time. The diffusion coefficient (equation~(\ref{eq:diffusion-coef-tumble})) to lowest order is equal to zero for tumble-time zero and reaches a maximum asymptotic value $Q_{xv}\tau_\tumble^2$ for very long tumble-times ($t_\tumble\gg\tau_\tumble$) as it is shown in Figure~\ref{fig:MSD-tumble}. Besides, the square ballistic coefficient is equal to $v_0=\sqrt{\sigma\,Q_{xv}\tau_\tumble}$ at tumble-time zero and converges to zero for very long tumble-times as shown in Figure~\ref{fig:MSD-tumble}. In addition, Figure~\ref{fig:MSD-tumble} shows that the ballistic behaviour is more important than the diffusion for $t_\tumble\lessapprox\tau_\tumble$. Left plot of Figure~\ref{fig:MSD-tumble} shows a global maximum of ballistic contribution at $t^*=0$ for $2/3\le\sigma\le 1$. Right plot of Figure~\ref{fig:MSD-tumble} shows a local minimum at $t^*=0$ and a global maximum at $0< t^*\le\tau_\tumble\,\ln 2$ for $0\le\sigma <2/3$. Considering equations~(\ref{eq:diffusion-coef-tumble}), (\ref{eq:ballistic-coef-tumble}) and (\ref{eq:v-corr}), the mean square velocity (MSV) at tumble-times yields 
\begin{equation}
\bigl\langle [v(t_\tumble)]^2\bigr\rangle=\frac{1}{\tau_\tumble}\,D(t_\tumble)+B(t_\tumble)\;.
\end{equation}
Assuming an exponential distribution of tumble-times $P(t_\tumble)=(\lambda_\tumble/\tau_\tumble)\,\ex^{-\lambda_\tumble t_\tumble/\tau_\tumble}$, the average of the square ballistic coefficient is
\begin{equation}
\bar{B}_\tumble=Q_\mathrm{vv}\tau_\tumble\,\frac{(2+\sigma\lambda_\tumble)\,\lambda_\tumble}{(\lambda_\tumble+1)(\lambda_\tumble+2)}\;
\end{equation}
and the average of the diffusion coefficient is $\bar{D}_\tumble=\bar{B}_\tumble\,\bar{t}_\tumble\,$, where $\bar{t}_\tumble=\tau_\tumble/\lambda_\tumble$ is the mean tumble-time. Experiments show that  $\lambda_\tumble\approx 1$ \cite{Berg1972}; then, taking $\lambda_\tumble = 1$ we find that $\frac{1}{3}\le\bar{B}_\tumble/(Q_\mathrm{vv}\tau_\tumble)\le\frac{1}{2}$ if $0\le\sigma\le 1$. 

\section{Statistical properties of the run motion\label{sec:four}}

One statistic observable of interest is the directional correlation function of the run motion. Making use of equation~(\ref{eq:e(t)}) and noting that the angle between two directions is $\theta(t,t')=\varphi(t)-\varphi(t')$, it turns out for small deflections that
\begin{equation}
\bigl\langle \mathbf{e}(t)\cdot\mathbf{e}(t')\bigr\rangle = \bigl\langle \cos[\theta(t,t')]\bigr\rangle \,\approxeq\, \bigl\langle x(t) x(t')\bigr\rangle\;.\label{eq:dir-correlation-run}
\end{equation}
Taking into account that $x_\mathrm{s}(\beta_\run)=1$ (equation~(\ref{eq:sl-x_st})), the deflection correlation function is
\begin{equation}
\bigl\langle x(t_1)\,x(t_2)\bigr\rangle = 1+b_x\Bigl(\ex^{-t_1/\tau_\run}+\ex^{-t_2/\tau_\run}\Bigr)+ b_x^2\,\ex^{-(t_1+t_2)/\tau_\run}+\epsilon_x\Bigl[\ex^{-\abs{t_1-t_2}/\tau_\run}-\ex^{-(t_1+t_2)/\tau_\run}\Bigr]\,,\label{eq:x-corr-run}
\end{equation}           
where \mbox{$b_x =x_0-1$} and \mbox{$\epsilon_x =Q_{xx}\,\tau_\run\,$}, with \mbox{$0\lessapprox \abs{b_x}\le \epsilon_x$} and $b_x\lessapprox 0$ considering the initial condition $x_0\lessapprox 1$. The directional correlation function satisfies \mbox{$\bigl\langle \mathbf{e}(t_1)\cdot\mathbf{e}(t_2)\bigr\rangle\leq 1$}. Consequently, the correlation function satisfies $\bigl\langle x(t_1)\,x(t_2)\bigr\rangle\lessapprox 1$ for small deflections. If the initial condition is very close to the stable state solution, {\sl i.e.} $\abs{b_x}\ll 1$, using equation~(\ref{eq:x-corr-run}) we find the inequality
\begin{equation}
g(t,T)=\frac{1+\ex^{T/\tau_\run}}{2\,\sinh(t/\tau_\run)}\;\gtrapprox\; \frac{\epsilon_x}{\abs{b_x}}=\frac{Q_{xx}\tau_\run}{1-x_0}\;,\label{eq:ineq-run}
\end{equation}
where $t=\min(t_1,t_2)$ and $T=\abs{t_1-t_2}$. The plot of Figure 3 shows $g$ as a function of time $t$ for several values of $T$. We observe that there is a minimum time $t_{\min}\le\tau_\run\,\arcsinh[{\abs{b_x}}/(2\,\epsilon_x)]$, defined by the autocorrelation ($T=0$), time from which the bacterium stops its runs. In fact, the experimental observations confirm the existence of a minimum runtime \cite{Berg1972}, which based on equation~(\ref{eq:ineq-run}) can be related to the noise intensity and the initial condition of each run.
\begin{figure}[h!]
\centering
\includegraphics[scale=0.55]{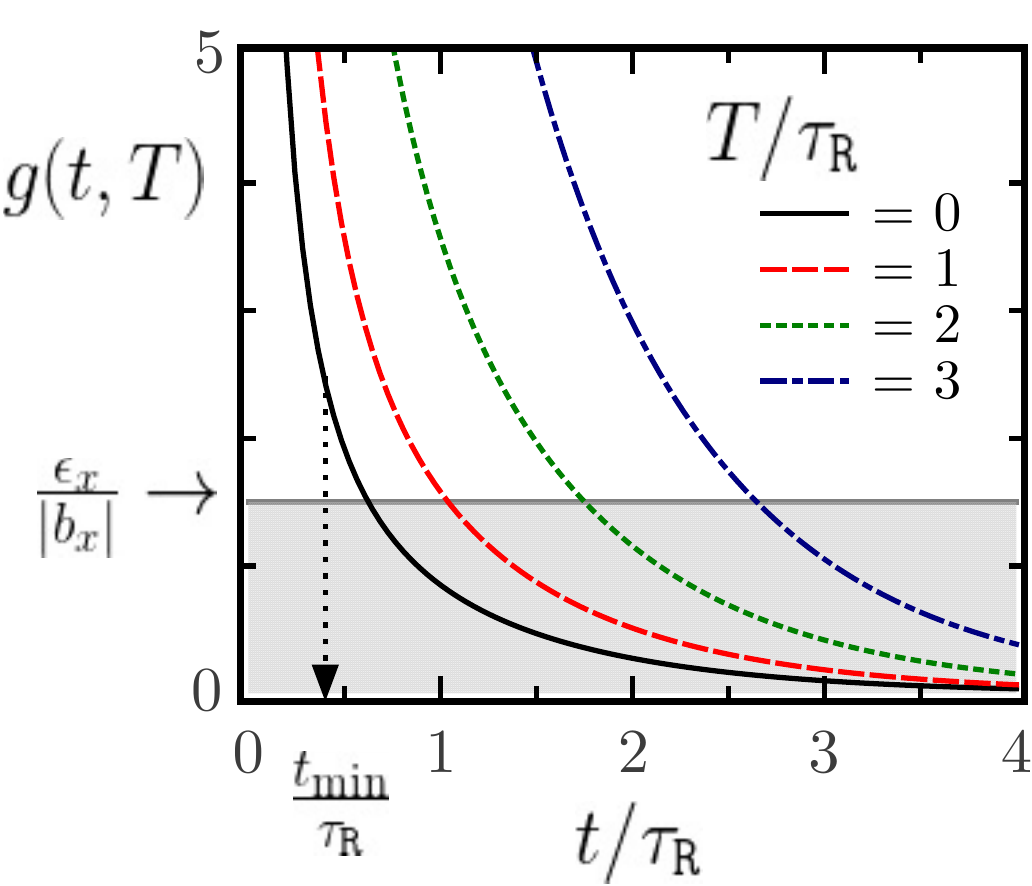}
\caption{(color online) Shows $g(t,T)$ as a function of time $t$ for several values of $T$ (see left-side of the inequality given by equation~(\ref{eq:ineq-run})). In the non-shaded region $g\geqslant \epsilon_x/\abs{b_x}$. The minimum runtime is defined in the interval $0< t_{\min}\le\tau_\run\,\arcsinh[{\abs{b_x}}/(2\,\epsilon_x)]$, as the plot shows. 
\label{fig:DCF-run}}
\end{figure}
Taking into account that $v_\mathrm{s}(\beta_\run)=1$ (equation~(\ref{eq:vs-beta})), the speed correlation function is
\begin{equation}
\bigl\langle v(t_1)\,v(t_2)\bigr\rangle = 1+b_v\Bigl(\ex^{-t_1/\tau_\run}+\ex^{-t_2/\tau_\run}\Bigr)+ b_v^2\,\ex^{-(t_1+t_2)/\tau_\run}+\epsilon_v\Bigl[\ex^{-\abs{t_1-t_2}/\tau_\run}-\ex^{-(t_1+t_2)/\tau_\run}\Bigr]\,,\label{eq:v-corr-run}
\end{equation}           
where \mbox{$b_v =v_0-1$} and \mbox{$\epsilon_v =Q_{vv}\,\tau_\run\,$}, with $0\lessapprox \abs{b_v}\le \epsilon_v$\,. We assume the initial condition \mbox{$v_0=v_\mathrm{s}(1+b_v)$} with \mbox{$\abs{b_v}\ll 1$}, so that \mbox{$v_0\approx 1$}. Equations~(\ref{eq:x-corr-run}) and (\ref{eq:v-corr-run}) are useful in order to determine the MSD of the run motion assuming that the processes $(x,v)$ are independent, which is equivalent to taking $Q_{xv} = 0$, or
\begin{equation}
\bigl\langle\mathbf{v}(t_1) \cdot\mathbf{v}(t_2)\bigr\rangle= \bigl\langle v(t_1)\,v(t_2)\bigr\rangle\;\bigl\langle \mathbf{e}(t_1)\cdot\mathbf{e}(t_2)\bigr\rangle\;,
\end{equation}
where the directional correlation function can be approximated by equation~(\ref{eq:dir-correlation-run}) assuming small deflections. The assumption that the processes $(x, v)$ are independent is in agreement with the experimental fact that there is no flagellar unbundling during the run motion and, consequently, there is no reason for a net displacement perpendicular to the translation direction.
\begin{figure}[h!]
\centering
\includegraphics[scale=0.5]{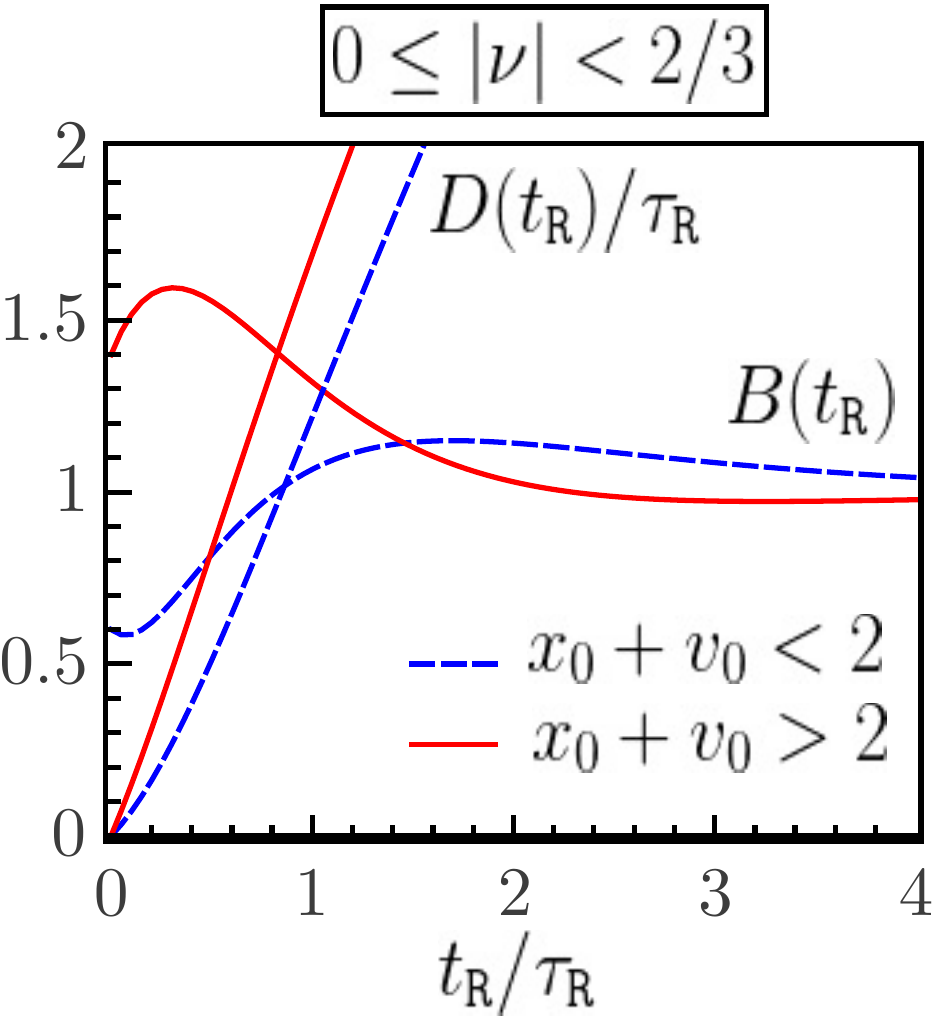}\quad
\includegraphics[scale=0.5]{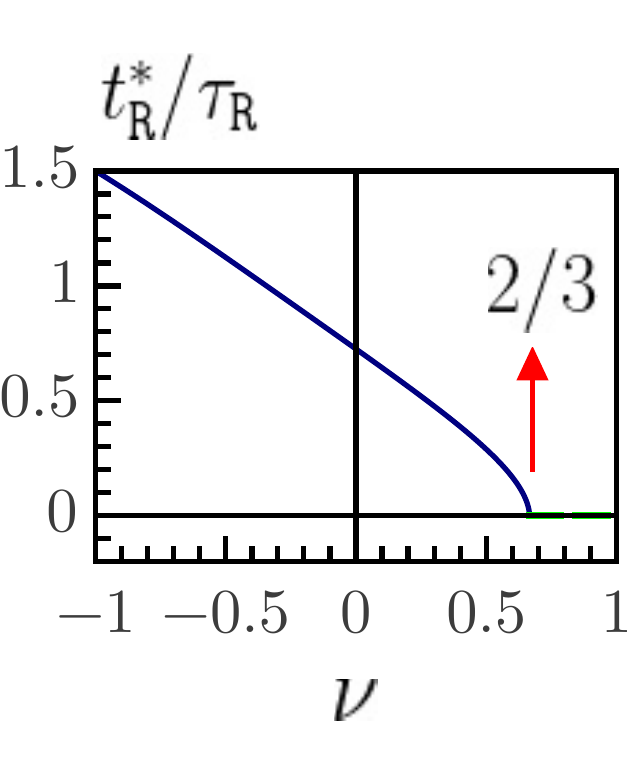}
\includegraphics[scale=0.5]{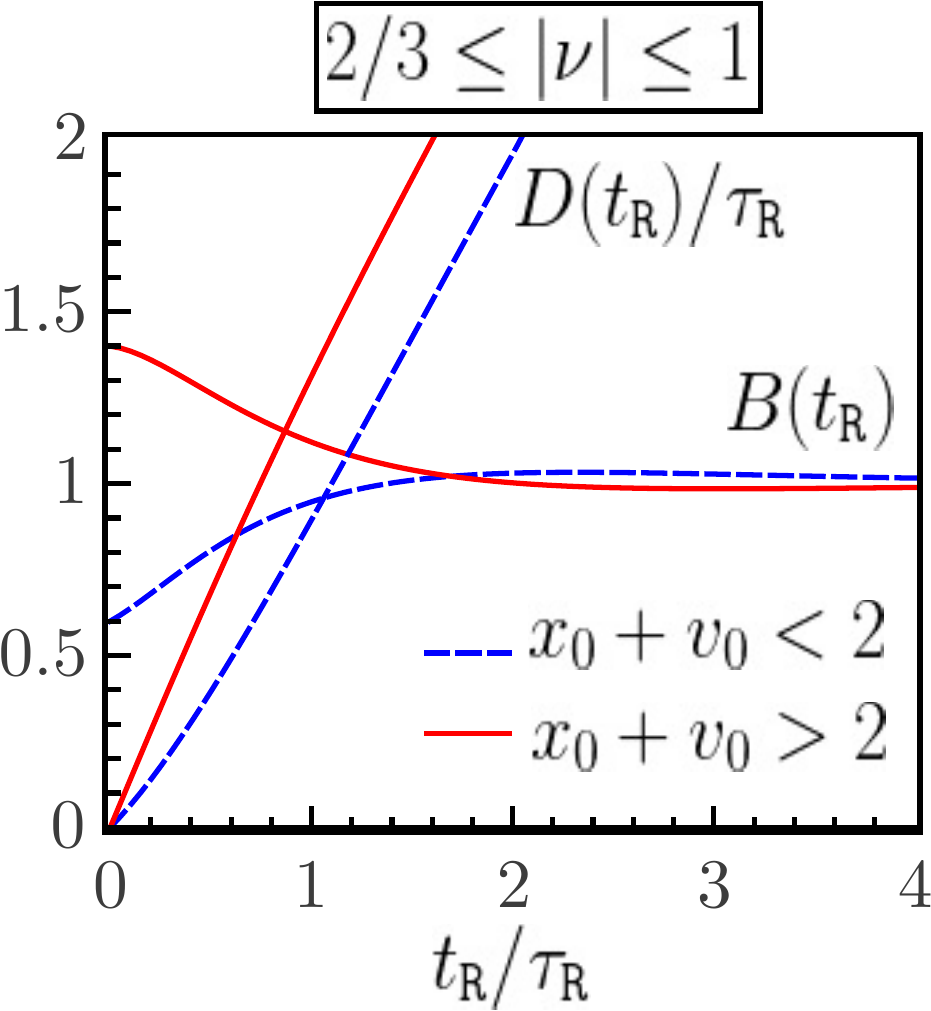}
\caption{(color online) Left and right plots: show the diffusion coefficient $D(t_\run)$ and square ballistic coefficient $D(t_\run)$ as a function of the runtime $t_\run$ for two different intervals of the parameter $\nu$ (defined by equation~(\ref{eq:nu})). Both functions and time are dimensionless. These plots were made taking \mbox{$x_0+v_0 =2\pm 0.2$} (both plots), \mbox{$\epsilon_x=\epsilon_v= 1/2$} (left plot) and \mbox{$\epsilon_x=\epsilon_v = 1/8$} (right plot). Middle plot: shows the adimensionalized runtime $t^*_\run$ as a function of $\nu$, where $t^*_\run$ is the time where $B(t_\run)$ reaches a global maximum. This plot was made taking \mbox{$\epsilon_x=\epsilon_v= 1/2$}.\label{fig:MSD-run}}
\end{figure}
The MSD for the run motion is
\begin{flalign}
\RMSD (t) = \tau_\run^2\,\biggl\{&\Bigl(\frac{t}{\tau_\run}\Bigr)^{\!2}+2\,(b_x+b_v)\,\Bigl(\frac{t}{\tau_\run}\Bigr)\Bigl(1-\ex^{-t/\tau_\run}\Bigr)\nonumber\\ 
&+(\epsilon_x+\epsilon_v)\Bigr(-3+2\,\frac{t}{\tau_\tumble}+4\,\ex^{-t/\tau_\run}-\ex^{-2\,t/\tau_\run}\Bigr)\nonumber\\
&+(\epsilon_x\,b_v+\epsilon_v\,b_x)\Bigl[2-\Bigl(3+2\,\frac{t}{\tau_\run}\Bigr)\ex^{-t/\tau_\run}+2\,\ex^{-2\,t/\tau_\run}-\ex^{-3\,t/\tau_\run}\Bigr]\nonumber\\
&+\epsilon_x\,\epsilon_v\,\Bigl[-\frac{5}{4}+\frac{t}{\tau_\run}+\Bigl(1+2\,\frac{t}{\tau_\run}\Bigr)\ex^{-2\,t/\tau_\run}+\frac{1}{4}\,\ex^{-4\,t/\tau_\run}\Bigr]+\Or(b^2)\biggr\}\;,
\end{flalign}
where $\Or(b^2)$ contains terms whose order is greater than $1$ in the parameters $b_x$ and $b_v$\,. In analogy to the tumble-motion, the usual asymptotic analysis to determine the diffusion constant is meaningless if the bacterium stops its run motion at runtimes whose order is that of the characteristic runtime. In consequence, to study the MSD behaviour, we calculate its Taylor series at times close to the runtime ({\sl i.e.} $t\lessapprox t_\tumble$)
\begin{equation}
\RMSD(t)=\RMSD(t_\run)+2\,D(t_\run)\,(t-t_\run)+B(t_\run)\,(t-t_\run)^2+\cdots\;,\label{eq:MSD-Taylor-run}
\end{equation}
where 
\begin{flalign}
D(t_\run)&=\frac{1}{2}\;\frac{\upd\RMSD}{\upd\,t}\biggr\rfloor_{t=t_\run}&\nonumber\\
&=\tau_\run\,\biggl\{\frac{t_\run}{\tau_\run}+ (b_x+b_v)\,\Bigl[1-\Bigl(1-\frac{t_\run}{\tau_\run}\Bigr)\,\ex^{-t_\run/\tau_\run}\Bigr]+(\epsilon_x +\epsilon_v)\,\Bigr(1-\ex^{-t_\run/\tau_\run}\Bigl)^{\!2}\nonumber\\
&\hspace{1.6cm}+(\epsilon_x\,b_v+\epsilon_v\,b_x)\Bigl(\frac{1}{2}+\frac{t_\run}{\tau_\run}-2\,\ex^{-t_\run/\tau_\run}+\frac{3}{2}\,\ex^{-2\,t_\run/\tau_\run}\Bigr)\,\ex^{-t_\run/\tau_\run}\nonumber\\
&\hspace{1.6cm}+\epsilon_x\,\epsilon_v\,\Bigl(\frac{1}{2}-2\,\frac{t_\run}{\tau_\run}\,\ex^{-2\,t_\run/\tau_\run}-\frac{1}{2}\,\ex^{-4\,t_\run/\tau_\run}\Bigr)+\Or(b^2)\biggr\}\label{eq:diffusion-coef-run}
\end{flalign}
is the diffusion coeficient at the runtime and 
\begin{flalign}
B(t_\run)&=\frac{1}{2}\;\frac{\upd^2 \RMSD}{\upd\,t^2}\biggr\rfloor_{t=t_\run}&&\nonumber\\
&=1+ (b_x+b_v)\,\Bigl(2-\frac{t_\run}{\tau_\run}\Bigr)\,\ex^{-t_\run/\tau_\run}+2\,(\epsilon_x+\epsilon_v)\Bigl(1-\ex^{-t_\run/\tau_\run}\Bigr)\,\ex^{-t_\run/\tau_\run}&\nonumber\\
&\hspace{0.7cm}+(\epsilon_x\,b_v+\epsilon_v\,b_x)\Bigl(\frac{1}{2}-\frac{t_\run}{\tau_\run}+4\,\ex^{-t_\run/\tau_\run}-\frac{9}{2}\,\ex^{-2\,t_\run/\tau_\run}\Bigr)\,\ex^{-t_\run/\tau_\run}\nonumber\\
&\hspace{0.7cm}+2\,\epsilon_x\,\epsilon_v\,\Bigl(-1+2\,\frac{t_\run}{\tau_\run}+\ex^{-2\,t_\run/\tau_\run}\Bigr)\,\ex^{-2\,t_\run/\tau_\run}+\Or(b^2)\label{eq:ballistic-coef-run}
\end{flalign}
is the square ballistic coefficient at the runtime. If the initial conditions satisfy \mbox{$v_0+x_0\gtrapprox 2$} or \mbox{$b_v+b_x\gtrapprox 0$} (see red-solid line plots of Figure~\ref{fig:MSD-run}) the ballistic contribution $B(t_\run)$ reaches a global maximum for runtimes \mbox{$t^*_\run > 0$} if $0\le\nu\le 2/3$ or \mbox{$t_\run^*=0$} if $2/3\le\nu\le 1$, where
\begin{equation}
\nu\,\approxeq\,\frac{b_x+b_v}{\epsilon_x+\epsilon_v}=\frac{x_0+v_0-2}{(Q_{xx}+Q_{vv})\tau_\run}\;.\label{eq:nu}
\end{equation}
This behaviour is similar to the one observed for the ballistic contribution in the tumble movement. In contrast, for \mbox{$v_0+x_0\lessapprox 2$} or \mbox{$b_v+b_x\lessapprox 0$} (see blue-dashed line plots of Figure~\ref{fig:MSD-run}) the ballistic contribution $B(t_\run)$ reaches a global minimum for runtimes \mbox{$t^*_\run\approx 0$} and reaches a global maximum for runtimes \mbox{$t^*_\run\approx \tau_\run$}. In both cases, $B(t_\run)\approxeq 1$ for long runtimes $t_\run\gg\tau_\run$. The diffusion coefficient shows common features independently of the initial conditions, as can be seen from left and right plots of Figure~(\ref{fig:MSD-run}). In all cases, \mbox{$D(t_\run)\approx [1+2\,(b_x+b_v)]\,t_\run$} for short runtimes $t_\run\ll\tau_\run$ and $D(t_\run)\approx t_\run$ for long runtimes $t_\run\gg\tau_\run$. The diffusion and square ballistic coefficients given by equations~(\ref{eq:diffusion-coef-run}) and (\ref{eq:ballistic-coef-run}), respectively, depend on $4$ parameters; this fact makes their analysis difficult. However, Figure~(\ref{fig:MSD-run}) shows the behaviour of the two remaining parameters for the special case of equal noise intensities, {\sl i.e.} $Q_{xx}=Q_{vv}$. Assuming an exponential distribution of runtimes $P(t_\run)=(\lambda_\run/\tau_\run)\,\ex^{-\lambda_\run t_\run/\tau_\run}$, the average of the square ballistic coefficient is
\begin{equation}
\bar{B}_\run=1+(b_x+b_v)\,\frac{(1+2\lambda_\run)\,\lambda_\run}{(\lambda_\run+1)^2}+
2\,(\epsilon_x+\epsilon_v)\frac{\lambda_\run}{(\lambda_\run+1)(\lambda_\run+2)}+8\,\epsilon_x\,\epsilon_v\frac{\lambda_\run}{(\lambda_\run+2)^2(\lambda_\run+4)}\;
\end{equation}
and the average of the diffusion coefficient is $\bar{D}_\run=\bar{B}_\run\,\bar{t}_\run\,$, where $\bar{t}_\run=\tau_\run/\lambda_\run$ is the mean runtime.

\section{Conclusions}

Previous to this work, the movements of run and tumble have been studied separately or as sequences of both movements. These studies have been primarily focused on a single run or sequence of runs abruptly interrupted by tumbles \cite{Condat2005}. Little attention has been paid to the theoretical modeling of the movement of the tumble, especially because of the limited availability of experimental data. Nevertheless, there are experimental data provided by the pioneering work of Berg and Brown \cite{Berg1972}. In their work statistical quantities for both movements were measured, such as the distributions of the tumble-angles and tumble-durations, and the ratio between the mean durations of the run and tumble. With these data, in a previous work \cite{Fier2017}, we were able to derive a single Langevin equation for the change of orientation or deflection $x(t)$ of the bacterium in the run and tumble movements. Additionally, to complete the theoretical model, in this paper we establish a unique Langevin equation for the speed $v(t)$ of the run and tumble movements. Each type of movement is characterized by values taken by a control parameter $\beta$. In particular, the steady state solutions as well as the characteristic times are functions of this parameter. Langevin equations are solved analytically, which makes it possible to calculate the statistical properties of each movement in detail. Assuming that the stochastic processes $(x,v)$ are not independent during the tumble, we show that there are small displacements of the center of mass of the bacterium in normal direction to the body axis of the bacterium. This result is in agreement with the observation of the flagellar unbundling during the CCW to CW transition. In addition, we show that the directional correlation during the tumble has non-stationary terms at tumble-times close to the characteristic time of this movement. For the tumble movement, we also derive the mean square displacement (MSD) and, at times close to the tumble-time, we determine the diffusion coefficient and the square ballistic coefficient. At very small tumble-times compared to the characteristic time, we observe that the ballistic contribution is more important than the contribution of the diffusion, which can be concluded from the maximum it shows at initial times. On the contrary, at tumble-times much longer than the characteristic time the ballistic coefficient goes to zero and the diffusion coefficient saturates. The statistical properties of the run movement are studied following the same methodology. Furthermore, we focus on studies of the MSD. First, we establish the conditions that need to satisfy the parameters so that the directional correlation is well defined. Assuming that the stochastic processes $(x, v)$ are independent, we calculate the MSD. We show that its behaviour depends on the initial conditions of speed and deflection as well as on the noise intensities linked to the variables. In general, it can be observed that the ballistic contribution to short runtimes is always more important than the diffusion. This behaviour seems to be trivial but depends on the initial conditions as well as on extrema values (maximum or minimum) of the ballistic coefficient at short runtimes. At long runtimes, it can be observed that the diffusion coefficient goes as time and the ballistic coefficient approaches a constant. 

As a final conclusion and outlook, we believe that this work can serve as basis for research of other flagellated bacteria with different movements to those analyzed here, such as the {\it Vibro alginolyticus}, an uni-flagellate bacterium that resides in marine environment, showing sequences of forward-run, reverse, backward-run and flick motions \cite{Stocker2011, Xie2011}.

\section*{Appendix}

\paragraph{A.} The integral of second term of equation~(\ref{eq:appendix-1}) is
\begin{equation}
\int_{t_0}^t G_x^{[\beta_\tumble]}(t,s)\,G_v^{[\beta_\tumble]}(t,s)\,\upd s =
\tau_\tumble\,\beta_\tumble\,\ex^{-t/\tau_\tumble}\,\ln(1-\beta_\tumble\,\ex^{-t/\tau_\tumble})\,\bigl[I_{xv}(\beta_\tumble\,\ex^{-t/\tau_\tumble})-I_{xv}(\beta_\tumble\,\ex^{-t_0/\tau_\tumble})\bigr]\;,
\end{equation}
where 
\begin{equation}
I_{xv}(u)=-\int^u\frac{\upd z}{z^2\,\ln(1-z)}\;.
\end{equation}
After integrating the series expansion of the function $I_{xv}$ around $z=0$ we obtain equation~(\ref{eq:appendix-2}). 
\paragraph{B.} We calculate the directional correlation (see equation~(\ref{eq:2-correlation})) by means of the correlation
\begin{eqnarray}
\bigl\langle [x(t)-x_\mathrm{s}(\beta_\tumble)][x(t')-x_\mathrm{s}(\beta_\tumble)]\bigr\rangle \!&=&\! \bigl[\langle x(t)\rangle-x_\mathrm{s}(\beta_\tumble)\bigr]\,\bigl[\langle x(t')\rangle-x_\mathrm{s}(\beta_\tumble)\bigr]\nonumber\\&&+\;2\,Q_{xx}\int_{t_0}^{\min(t,t')} \!G_x^{[\beta_\tumble]}(t,s)\,G_x^{[\beta_\tumble]}(t',s)\,\upd s\;.
\end{eqnarray}
The integral of the second term is (see reference \cite{Fier2017} for details)
\begin{eqnarray}
\int_{t_0}^{\min(t,t')} \!G_x^{[\beta_\tumble]}(t,s)\,G_x^{[\beta_\tumble]}(t',s)\,\upd s \!&=&\! \tau_\tumble\,\ln(1-\beta_\tumble\,\ex^{-t/\tau_\tumble})\,\ln(1-\beta_\tumble\,\ex^{-t'/\tau_\tumble})\nonumber\\
&&\times\,\Bigl[I_{xx}\bigl(\beta_\tumble\,\ex^{-\min(t,t')/\tau_\tumble}\bigr)-I_{xx}
\bigl(\beta_\tumble\,\ex^{-t_0/\tau_\tumble}\bigr)\Bigr]\;,
\end{eqnarray}
where
\begin{equation}
I_{xx}(u)=-\int^u \frac{\upd z}{z\,\ln^2(1-z)}\;.
\end{equation}
After integrating the series expansion of the function $I_{xx}$ around $z=0$ we obtain equation~(\ref{eq:appendix-3}).

\section*{Acknowledgements}
This work was partially supported by Consejo Nacional de Investigaciones Cient{\'i}ficas y T{\'e}cnicas (CONICET), Argentina, PIP 2014/16 No. 112-201301-00629. R.C.B. thanks C. Rabini for her suggestions on the final manuscript.

\bibliography{BFH.model-biblio}

\begin{thebibliography}{10}
\providecommand{\url}[1]{\texttt{#1}}
\providecommand{\urlprefix}{URL }
\expandafter\ifx\csname urlstyle\endcsname\relax
  \providecommand{\doi}[1]{doi:\discretionary{}{}{}#1}\else
  \providecommand{\doi}{doi:\discretionary{}{}{}\begingroup
  \urlstyle{rm}\Url}\fi
\providecommand{\bibAnnoteFile}[1]{%
  \IfFileExists{#1}{\begin{quotation}\noindent\textsc{Key:} #1\\
  \textsc{Annotation:}\ \input{#1}\end{quotation}}{}}
\providecommand{\bibAnnote}[2]{%
  \begin{quotation}\noindent\textsc{Key:} #1\\
  \textsc{Annotation:}\ #2\end{quotation}}
\providecommand{\eprint}[2][]{\url{#2}}

\bibitem{Eisenbach2004}
Eisenbach M (2004) Chemotaxis.
\newblock Imperial College Press.
\bibAnnoteFile{Eisenbach2004}

\bibitem{Pedley1992}
Pedley TJ, Kessler JO (1992) Hydrodynamic phenomena in suspensions of swimming
  microorganims.
\newblock Annu Rev Fluid Meek 24: 313-358.
\bibAnnoteFile{Pedley1992}

\bibitem{Purcell1977}
Purcell EM (1977) Life at low {Reynolds} number.
\newblock American J of Phys 45: 3.
\bibAnnoteFile{Purcell1977}

\bibitem{Lauffenburger1991}
Lauffenburger DA (1991) Quantitative studies of bacterial chemotaxis and
  microbial-population dynamics.
\newblock Microb Ecol 22: 175-185.
\bibAnnoteFile{Lauffenburger1991}

\bibitem{Webre2003}
Webre DJ, Wolanin PM, Stock JB (2003) Bacterial chemotaxis.
\newblock Curr Biol 13: 47.
\bibAnnoteFile{Webre2003}

\bibitem{Berg1973}
Berg HC, Anderson RA (1973) Bacteria swim by rotating their flagellar
  filaments.
\newblock Nature 245: 380-382.
\bibAnnoteFile{Berg1973}

\bibitem{Macnab1977}
Macnab RM (1977) Bacterial flagella rotating in bundles: a study in helical
  geometry.
\newblock Proc Natl Acad Sci 74: 221-225.
\bibAnnoteFile{Macnab1977}

\bibitem{Turner2000}
Turner L, Ryu WS, Berg HC (2000) Real-time imaging of fluorescent flagellar
  filaments.
\newblock J Bacteriol 182: 2793-2801.
\bibAnnoteFile{Turner2000}

\bibitem{Darnton2007}
Darnton NC, Turner L, Rojevsky S, Berg HC (2007) On torque and tumbling in
  swimming \textit{{E}scherichia coli}.
\newblock J Bacteriol 189: 1756.
\bibAnnoteFile{Darnton2007}

\bibitem{Berg2004}
Berg HC (2004) {\it E. coli} in motion.
\newblock Spinger-Verlag.
\bibAnnoteFile{Berg2004}

\bibitem{Amselem2012}
Amselem G, Theves M, Bae A, Bodenschatz E, Beta C (2012) A stochastic
  description of \textit{{D}ictyostelium} chemotaxis.
\newblock PLoS ONE 7: e37213.
\bibAnnoteFile{Amselem2012}

\bibitem{Selmeczi2005}
Selmeczi D, Mosler S, Hagedorn PH, Larsen NB, Flyvbjerg H (2005) Cell motility
  as persistent random motion: Theories from experiments.
\newblock Biophys J 89: 912–931.
\bibAnnoteFile{Selmeczi2005}

\bibitem{Schienbein1993}
Schienbein M, Gruler H (1993) Langevin equation, {Fokker-Planck} equation and
  cell-migration.
\newblock Bull Math Biol 55: 585-608.
\bibAnnoteFile{Schienbein1993}

\bibitem{Selmeczi2007}
Selmeczi D, Tol\'{\i}c-N{\o}rrelykke SF, Sch\"aeffer E, Hagedorn PH, Mosler S,
  et~al. (2007) Springer-Verlag, volume 711 of \emph{Lecture Notes in Physics},
  chapter 9. Brownian Motion after Einstein: Some New Applications and New
  Experiments.
\newblock pp. 181-199.
\bibAnnoteFile{Selmeczi2007}

\bibitem{Romanczuk2012}
Romanczuk P, B\"ar M, Ebeling W, Lindner B, Schimansky-Geie L (2012) From
  individual to collective stochastic dynamics.
\newblock Eur Phys J Special Topics 202: 1-162.
\bibAnnoteFile{Romanczuk2012}

\bibitem{Fier2017}
Fier G, Hansmann D, Buceta RC (2017) A stochastic model for directional changes
  of swimming bacteria.
\newblock Soft Matter 13: 3385-3394.
\bibAnnoteFile{Fier2017}

\bibitem{Berg1972}
Berg HC, Brown DA (1972) Chemotaxis in \textit{{E}scherichia coli} analysed by
  three-dimensional tracking.
\newblock Nature 239: 500.
\bibAnnoteFile{Berg1972}

\bibitem{Saragosti2012}
Saragosti J, Silberzan P, Buguin A (2012) Modeling \textit{{E}. coli} tumbles
  by rotational diffusion. implications for chemotaxis.
\newblock PLoS ONE 7: e35412.
\bibAnnoteFile{Saragosti2012}

\bibitem{Schweitzer2003}
Schweitzer F (2003) Brownian Agents and Active Particles: {Collective} Dynamics
  in the Natural and Social Sciences.
\newblock Springer-Verlag.
\bibAnnoteFile{Schweitzer2003}

\bibitem{Gruler1994}
Gruler H, d~Boisfleury-Chevance A (1994) Directed cell movement and cluster
  formation: {Physical principles.}
\newblock J Physique I (France) 4: 1085-1105.
\bibAnnoteFile{Gruler1994}

\bibitem{Doi2001}
Doi M, Edwards SF (2001) The theory of polymer dynamics.
\newblock Oxford University Press.
\bibAnnoteFile{Doi2001}

\bibitem{Risken1989}
Risken H (1989) The Fokker-Planck Equation: Methods of Solution and
  Applications.
\newblock Springer-Verlag.
\bibAnnoteFile{Risken1989}

\bibitem{Condat2005}
Condat CA, J\"ackle J, Mench\'on SA (2005) Randomly curved runs interrupted by
  tumbling: A model for bacterial motion.
\newblock Phys Rev E 72: 021909.
\bibAnnoteFile{Condat2005}

\bibitem{Stocker2011}
Stocker R (2011) Reverse and flick: Hybrid locomotion in bacteria.
\newblock PNAS 108: 2635-2636.
\bibAnnoteFile{Stocker2011}

\bibitem{Xie2011}
Xie L, Altindal T, Chattopadhyay S, Wu XL (2011) Bacterial flagellum as a
  propeller and as a rudder for efficient chemotaxis.
\newblock PNAS 108: 2246-2251.
\bibAnnoteFile{Xie2011}

\end{thebibliography}
\end{document}